# MIMO APP Receiver Processing with Performance-Determined Complexity

Konstantinos Nikitopoulos, *Member, IEEE,* and Gerd Ascheid,

*Senior Member, IEEE*

## Abstract

Typical receiver processing, targeting always the best achievable bit error rate performance, can result in a waste of resources, especially, when the transmission conditions are such that the best performance is orders of magnitude better than the required. In this work, a processing framework is proposed which allows adjusting the processing requirements to the transmission conditions and the required bit error rate. It applies *a-posteriori* probability receivers operating over multiple-input multiple-output channels. It is demonstrated that significant complexity savings can be achieved both at the soft, sphere-decoder based detector and the channel decoder with only minor modifications.

## Index Terms

MIMO systems, soft-output detection, sphere decoding

## I. INTRODUCTION

Multiple-input, multiple-output (MIMO) transmission offers increased spectral efficiency by concurrently transmitting multiple streams over the same frequency band. Therefore, MIMO systems have been adopted by several upcoming wireless communication standards like IEEE 802.11n and IEEE 802.16e. A plethora of approaches have been proposed for the detection and decoding of such systems providing different complexity/performance tradeoffs. The linear

The authors are with the Faculty of Electrical Engineering and Information Technology, RWTH Aachen University, Aachen, Germany (e-mail: Konstantinos.Nikitopoulos@iss.rwth-aachen.de).

This work has been supported by the UMIC Research Center, RWTH Aachen University and by the EU FP7 NEWCOM++ (Grant No. 216715) project.

detection methods (e.g., zero-forcing, MMSE), which calculate hard outputs and employ hard channel decoding are of low complexity but provide reduced performance. Conversely, the *a-posteriori* probability (APP) receivers provide increased performance at the cost of highly increased processing requirements [1].

Although APP receiver processing may be demanded for achieving the required error-rate performance over "unfavorable" transmission environments (i.e., over ill-conditioned MIMO transmission channels and at a low signal-to-noise ratio (SNR) regime) it may result in a significant waste of resources under good transmission scenarios where even less complex solutions could provide the required performance. Such a waste of resources, in terms of energy consumption and latency, is unavoidable for fixed algorithmic solutions targeting the "worst-case" scenario.

In order to avoid unnecessary complexity, multi-modal receivers employing different detection/decoding approaches with respect to the transmission scenario have been proposed [2], [3]. However, such approaches introduce an area overhead which is a function of the supported algorithmic approaches (i.e., modes). In addition, they require computationally intensive selection strategies for choosing the less complex supported mode from the set of the available ones who can provide the target bit error rate (TER) performance. These selection strategies involve *performance prediction* methods for any of the supported modes and have to be repeated any time a performance-affecting parameter changes (e.g., the specific channel realization). Therefore, the applicability of multi-modal receivers is restricted to those scenarios where the performance prediction is both available (i.e., analytical methods for predicting performance exist) and feasible (i.e., of low processing overhead).

The proposed approach assumes a single soft-output detector able to adjust its processing requirements to the transmission conditions and the TER. It is based on the depth-first sphere decoder (SD) of [1] and it can provide detection performance from *max-log* MAP down to the one of order successive interference cancelation [4]. In contrast to the aforementioned multi-modal approaches, the proposed one provides significant complexity reduction but it *does not* target the minimum detection complexity. Minimizing detector's complexity would require the exploitation of the full correcting capabilities of the channel code in order to, finally, reach the TER. Therefore, it would prevent from complexity savings at the channel decoder side.

In the proposed approach, the detection processing requirements are reduced by approximately calculating the soft information of those bits which already meet the TER requirements before channel decoding. As it will be explained in detail, such an approximation is feasible since it does not significantly affect the performance of the channel decoder. Then, when systematic codes are used, selective channel decoding can be performed only for the bits which have not already reached the TER before decoding (and their soft information has not been approximated) resulting in additional savings at the channel decoder side. Therefore, the overall complexity gains of the receiver are finally distributed to both the detector and the channel decoder. In addition, the proposed approach avoids the *performance prediction* burden which is required to minimize the detector's complexity for a given code and a given reduced complexity detector. In this way, increased practicality and applicability is achieved.

In order to adjust the SD complexity to the transmission conditions and the TER, the well-known concept of log-likelihood ratio (LLR) clipping is employed. This kind of clipping bounds the dynamic range of the LLRs and reduces the detector's complexity at the cost of reduced performance. The concept of LLR clipping has been originally proposed in [1] in order to align the detector's complexity to the unavoidable performance loss originating from the fixed-point implementation. Therefore, the LLR clipping value was determined by the selected fixed-point accuracy (typically selected via extensive simulations which link the fixed-point accuracy to the achievable performance). In this work, the concept is extended in the context of scenario-adaptive receiver processing. A simple and practical *performance driven* LLR clipping is proposed in order to choose the clipping value "on-demand", according to the TER performance. Adjusting the receiver's complexity by changing LLR the clipping value does not introduce any significant processing overhead since, typically, clipping is inherent in the depth-first SD approaches similar to [1]. It is noted that the performance driven LLR clipping still incorporates the ability to align the detector's complexity to the selected fixed-point accuracy when the accuracy is linked to the related bit error-rate performance.

## II. APP RECEIVER PROCESSING FOR MIMO SYSTEMS

The soft-output detector operates over several MIMO channel utilizations. It employs the received vectors $\mathbf{y}$ in order to calculate the *a-posteriori* soft information $\mathbf{L}_D$ of the coded bits. The resulting soft information is de-interleaved and fed to the soft-input, soft-output (SISO) channel decoder as *a-priori* information $\tilde{\mathbf{L}}_A$ in order to calculate channel decoder's *a-posteriori* soft information $\tilde{\mathbf{L}}_D$. Finally, the decoded bits are calculated from the sign of $\tilde{\mathbf{L}}_D$.

## A. Soft Detection in Terms of Sphere Decoding

In MIMO transmission with $M_T$ transmit and $M_R \geq M_T$ receive antennas, at the $u$-th MIMO channel utilization, the interleaved coded bits are grouped into blocks $B_{t,u}$ ($t = 1, ..., M_T$ and $u = 1, ..., U$ with $U$ being the number of channel utilizations per code block) in order to be mapped onto symbols $s_{t,u}$ of a constellation set $S$ of cardinality $|S|$. The bipolar $k$-th bit resides in block $B_{\lceil k/\log_2 |S| \rceil, u}$ and the blocks $B_{t,u}$ are mapped onto the symbols $s_{t,u}$ by a given mapping function (e.g., Gray mapping). The corresponding received $M_R \times 1$ vector $\mathbf{y}_u$ is, then, given by

$$\mathbf{y}_u = \mathbf{H}_u \mathbf{s}_u + \mathbf{n}_u, \tag{1}$$

where $\mathbf{H}_u$ is the $M_R \times M_T$ complex channel matrix and $\mathbf{s}_u = [s_{1,u}, s_{2,u}, ..., s_{M_T,u}]^T$ is the transmitted symbol vector. Then, $c_{b,i,u}$ is the $b$-th bit of the $i$-th entry of $\mathbf{s}_u$ and the term $\mathbf{n}_u$ is the noise vector, consisting of i.i.d., zero-mean, complex, Gaussian samples with variance $2\sigma_n^2$.

The soft-output detector calculates the *a-posteriori* LLRs for all the symbols residing in the frame to be decoded. Namely, it calculates

$$L_D\left(c_{b,i,u}\right) = \ln\left(\frac{P[c_{b,i,u} = +1 | \mathbf{y}_u, \mathbf{H}_u]}{P[c_{b,i,u} = -1 | \mathbf{y}_u, \mathbf{H}_u]}\right), \ \forall b, i, u. \tag{2}$$

Assuming that the corresponding bits are statistically independent (due to interleaving) and under the *max-log* approximation, the problem transforms to

$$L_D\left(c_{b,i,u}\right) \approx \frac{1}{2\sigma_n^2} \min_{\mathbf{s}_u \in S_{b,i,u}^{-1}} \|\mathbf{y}_u - \mathbf{H}_u \mathbf{s}_u\|^2 - $$
$$\frac{1}{2\sigma_n^2} \min_{\mathbf{s}_u \in S_{b,i,u}^{+1}} \|\mathbf{y}_u - \mathbf{H}_u \mathbf{s}_u\|^2 \tag{3}$$

where $S_{b,i,u}^{\pm 1}$ are the sub-sets of possible $\mathbf{s}_u$ symbol sequences having the $b$-th bit value of their $i$-th $\mathbf{s}_u$ entry equal to $\pm 1$.

In order to avoid exhaustive search the problem can be transformed into an equivalent tree-search which can be efficiently solved in terms of sphere decoding [5]. In detail, the channel matrix $\mathbf{H}_u$ can be QR decomposed into $\mathbf{H}_u = \mathbf{Q}_u \mathbf{R}_u$, with $\mathbf{Q}_u$ a unitary $M_R \times M_T$ matrix and $\mathbf{R}_u$ an $M_T \times M_T$ upper triangular matrix with elements $R_{i,j,u}$ and real-valued positive diagonal entries. Then, under the LLR calculation, (3) transforms to [1]

$$L_D\left(c_{b,i,u}\right) \approx \frac{1}{2\sigma_n^2} \min_{\mathbf{s}_u \in S_{b,i,u}^{-1}} \|\mathbf{y}'_u - \mathbf{R}_u \mathbf{s}_u\|^2 - $$

$$\frac{1}{2\sigma_n^2} \min_{\mathbf{s}_u \in S_{b,i,u}^{+1}} \|\mathbf{y'}_u - \mathbf{R}_u \mathbf{s}_u\|^2 \qquad (4)$$

where $\mathbf{y'}_u = \mathbf{Q}_u{}^H \mathbf{y}_u = \left[ y'_{1,u}, y'_{2,u}, ..., y'_{M_T,u} \right]^T$.

### B. SISO Channel Decoding

Similarly to the soft-output detector, after de-interleaving, the resulting soft-information is employed to calculate the corresponding *a-posteriori* information, as

$$\tilde{L}_D\left(\tilde{c}_k\right) = \ln\left( \frac{P[\tilde{c}_k = +1 | \tilde{\mathbf{L}}_A]}{P[\tilde{c}_k = -1 | \tilde{\mathbf{L}}_A]} \right) \qquad (5)$$

where for $\tilde{\mathbf{c}}$ being the encoded sequence after de-interleaving with elements $\tilde{c}_k$, (5) becomes

$$\tilde{L}_D\left(\tilde{c}_k\right) = \ln\left( \sum_{\tilde{\mathbf{c}}:\tilde{C}_k^{+1}} P\left[\tilde{\mathbf{c}}|\tilde{\mathbf{L}}_A\right] \right) - \ln\left( \sum_{\tilde{\mathbf{c}}:\tilde{C}_k^{-1}} P\left[\tilde{\mathbf{c}}|\tilde{\mathbf{L}}_A\right] \right) =$$

$$= \ln\left( \sum_{\tilde{\mathbf{c}}:\tilde{C}_k^{+1}} \exp\sum_{i=1}^{K} \ln P\left[\tilde{c}_i|\tilde{L}_A\left(\tilde{c}_i\right)\right] \right) -$$

$$\ln\left( \sum_{\tilde{\mathbf{c}}:\tilde{C}_k^{-1}} \exp\sum_{i=1}^{K} \ln P\left[\tilde{c}_i|\tilde{L}_A\left(\tilde{c}_i\right)\right] \right) \qquad (6)$$

with $\tilde{C}_k^{\pm 1}$ being the set of bit sequences $\tilde{\mathbf{c}}$, of length $K$, with their $k$-th bit equal to $\pm 1$. Then (6) can be efficiently calculated by the well-known BCJR-MAP algorithm [6], [7].

From (6) it becomes apparent that the most significantly contributing sequences $\tilde{\mathbf{c}}$ to the LLR calculation are those with their non-positive $\sum_{i=1}^{K} \ln P\left[\tilde{c}_i|\tilde{L}_A\left(\tilde{c}_i\right)\right]$ terms being close to zero. Therefore, these sequences do not contain highly unreliable bits of very low $P\left[\tilde{c}_i|\tilde{L}_A\left(\tilde{c}_i\right)\right]$, or equivalently, bits of high $\left|\tilde{L}_A\left(\tilde{c}_k\right)\right|$ value and sign opposite to $\tilde{L}_A\left(\tilde{c}_k\right)$'s. For this reason, saving complexity by employing approximate LLR calculation for the highly non-reliable bits is not expected to significantly affect the outcome of the channel decoder. Additionally, under the approximation of [8]

$$\ln P\left(\tilde{c}_k|\tilde{L}_A\left(\tilde{c}_k\right)\right) \approx \frac{1}{2}\left( \tilde{c}_k \tilde{L}_A\left(\tilde{c}_k\right) - \left|\tilde{L}_A\left(\tilde{c}_k\right)\right| \right) \qquad (7)$$

which holds for large $\left|\tilde{L}_A\left(\tilde{c}_k\right)\right|$ values (typically larger than 2) it can be deduced that for highly reliable bits (i.e., of high $\left|\tilde{L}_A\left(\tilde{c}_k\right)\right|$ value and sign equal to the one of $\tilde{L}_A\left(\tilde{c}_k\right)$) the terms $\ln P\left[\tilde{c}_i|\tilde{L}_A\left(\tilde{c}_i\right)\right]$ in (7) equal zero independently of the exact $\tilde{L}_A$ value. The last two observations

lead to the conclusion that approximate (and thus of lower complexity) calculation of the strong soft information, (i.e., of high $\left|\tilde{L}_A\left(\tilde{c}_k\right)\right|$) is not expected to significantly affect the outcome of the SISO channel decoder.

## III. Scenario-Adaptive MAP Receiver Processing

The proposed approach exploits the ability to calculate the probability of an erroneous hard decision of a specific bit by utilizing its soft information. Additionally, as it has been discussed in Section II.B, it benefits from the capability to approximate the strong soft information without noticeable consequences on the achievable performance. In detail, the proposed approach consists of the following steps.

1) *Linking the LLR values to the TER performance*

   According to [9] the error probability of the hard decision of the bit $c$ with *a-posteriori* LLR $L_c$ is

   $$P_e(c) = \frac{1}{1 + \exp\left(\left|L\left(c\right)\right|\right)}. \tag{8}$$

   Thus, the bit-error-rate (BER) of a code block can be approximated as

   $$\hat{P}_b \approx \frac{1}{N_I} \sum_{k=1}^{N_I} P_e\left(\tilde{c}_k^{(I)}\right) \tag{9}$$

   where $\hat{c}_i^{(I)}$ are the $N_I$ information bits. From the above equation it becomes apparent that the provided BER is dominated by the bits with small $\left|\tilde{L}_D\left(\tilde{c}_i^{(I)}\right)\right|$ values. Additionally, these are the bits which significantly contribute to the decoding process (see Eq. (6), Section II.B). Therefore, no approximation of these weak LLRs is attempted. From (9) it also becomes apparent that if all $P_e(\tilde{c}_k)$ values are lower then the TER, or equivalently, if $\left|\tilde{L}_D\left(\tilde{c}_k\right)\right| > \tilde{L}_{TER} = \ln(TER^{-1} - 1)$ for all bits, the average performance will also meet the TER. Based on the aforementioned observations unnecessary processing which will finally result in $\left|\tilde{L}_D\left(\tilde{c}_k\right)\right| > \tilde{L}_{TER}$ is avoided, while full processing takes place for the rest of the bits.

2) *Performing Scenario-Adaptive Soft-Output Detection*

   If the soft-output detector's *a-posteriori* information (which is the *a-priori* information for the SISO channel decoder) of a bit has met the TER constraint it is assumed that its LLR value is strong enough that employing an approximate LLR value will not significantly affect the outcome of the SISO channel decoder (see Section II.B). Thus, the complexity of the soft-output detector can be reduced by avoiding the extra processing dedicated to accurately calculate the soft information of the bits exceeding the TER constraint. This can be practically achieved by means of TER *performance-driven* LLR clipping or, in detail, by bounding the SD *a-posteriori* information in order not to exceed $\tilde{L}_{TER}$, as it has been explained at step 1.

The proposed approach allows approximations only for the bits which have already reached the TER before channel decoding. Then, for those specific bits, the *average* bit error rate performance after channel decoding is expected to be much better than the required TER. This observation leads to the discussion in Section I, according to which the proposed approach does not target the minimum SD complexity. If this was the case, the average error rate performance after channel decoding should just reach TER and it should not be better than that. Then, in order to minimize the decoding complexity tighter LLR clipping values should be set after *predicting* the performance gain provided by the channel decoder. This performance prediction would increase the computational burden and it would restrain the applicability and the practicality of the approach.

3) *Performing Scenario-Adaptive SISO (Systematic) Channel Decoding*

Following the same rationale described at steps 1 and 2, and for *systematic* channel codes, further complexity reduction can take place at the channel decoder side. In detail, *target-performance-driven, selective SISO decoding* can take place only for the bits which do not already meet the TER constrained before channel decoding (i.e., for the bits with $\left| \tilde{L}_A \left( \tilde{c}_k \right) \right| < \tilde{L}_{TER}$). For the rest hard decisions are taken based on the sign of their *a-priori* information. The way that the selective decoding can be translated to complexity savings, is discussed in Section III.B.

## A. Scenario-Adaptive Soft-Output Detection

Even if the proposed *performance-driven* LLR clipping under appropriate modifications is applicable to most of the SD approaches, the depth-first SD of [1] is herein assumed. In contrast to the *list* SD approaches of [5], [10]–[12], the SD of [1] can ensure the (exact) *max-log* MAP performance and, in addition, no modifications are required since the clipping procedure is already inherent. However, as also discussed in Section I, the LLR clipping has been originally proposed in order to adjust the detector's complexity to the fixed-point implementation. Therefore, in contrast to this work, no discussion has been made on how to choose the clipping value, or on how LLR clipping could be used for scenario-adaptive receiver processing.

The adopted SD employs depth-first tree traversal with Schnorr-Euchner enumeration and radius reduction (with infinite initial radius). In order to avoid those redundant calculations which are common to the different minimization problems (and tree searches) of (4) the single-tree-search approach of [1] is employed. According to this approach, only one tree search takes place but different radii are used for any of the minimization problems. LLR clipping is employed but, as already explained, instead of selecting the clipping value in accordance with the fixed-point implementation, *performance-driven* LLR clipping is employed with a clipping value of $\tilde{L}_{TER}$. More details about the SD structure and implementation can be found in [1].

## B. Scenario-Adaptive SISO (Systematic) Channel Decoder

As discussed in Section III, step 3, partial SISO channel decoding can be performed for systematic codes, only on the subset of bits not reaching the TER. For log-SISO approaches similar to [7], operating in the *log* domain and employing the $max^*$ operator in order to avoid the computational expensive multiplications, the expensive operations are not any more the corresponding calculations (which mainly become additions). Instead, as it is also shown in [13], [14], the expensive operations are the required energy consuming memory accesses and especially the ones related to the state metric storages. The significance effect of those memory accesses is also revealed in [13], [14], where in order to reduce them even additional extra processing is proposed. In the sequel, equivalently to [15], it is discussed how the selective LLR update may result in reduced number of state metric storages. However, this discussion is just indicative since the concept of selective SISO channel decoding cannot be quantified to energy savings without considering a specific implementation, which is beyond the scope of this work.

For an 1/2 convolutional code with $\tilde{c}_{x,t}(e)$ the encoder output bits for a transition $e$ from the state $s$ to $s'$ at coding time $t$ (with $x = 0, 1$) the corresponding $\tilde{L}_D(\tilde{c}_{x,t})$ can be expressed as [7]

$$\tilde{L}_D(\tilde{c}_{x,t}) = \max_{e:\tilde{c}_{x,t}=1} {}^* [\delta_t(e)] - \max_{e:\tilde{c}_{x,t}=-1} {}^* [\delta_t(e)] \tag{10}$$

with

$$\delta_t(e) = \alpha_{t-1}[s] + \tilde{c}_{0,t}(e)\tilde{L}_A(\tilde{c}_{0,t}(e)) + \tilde{c}_{1,t}(e)\tilde{L}_A(\tilde{c}_{1,t}(e)) + \beta_t[s'] \tag{11}$$

 

and $\alpha_t$, $\beta_t$ being the state metrics obtained through the following forward and backward recursions

$$\alpha_t(w) = \max_{e:s'=w} {}^* \left[ \alpha_{t-1}(s) + \tilde{c}_{0,t}(e)\tilde{L}_A\left(\tilde{c}_{0,t}(e)\right) + \tilde{c}_{1,t}(e)\tilde{L}_A\left(\tilde{c}_{1,t}(e)\right) \right] \tag{12}$$

$$\beta_t(w) = \max_{e:s=w} {}^* \left[ \beta_{t+1}(s') + \tilde{c}_{0,t+1}(e)\tilde{L}_A\left(\tilde{c}_{0,t+1}(e)\right) + \tilde{c}_{1,t+1}(e)\tilde{L}_A\left(\tilde{c}_{1,t+1}(e)\right) \right]. \tag{13}$$

As discussed in [14], the $\alpha$ values can be calculated and overwritten immediately as they are not required in future calculations. On the other hand, typically, all $\beta$ metrics need to be stored. However, for selective (per bit) channel decoding only the subset of $\beta$ values related to the decoded bits needs to be stored, resulting in potential energy consumption savings. Since, as we discussed, the energy savings can be quantified only for specific implementations, in Section IV we evaluate the potential gains in terms of required $\beta$ stores.

## IV. SIMULATIONS

A $4 \times 4$ MIMO system is assumed, operating over a spatially and temporally uncorrelated Rayleigh flat-fading channel. The encoded bits are mapped onto 16-QAM via Gray coding. A systematic $(5/7)_8$ recursive convolutional code of rate $1/2$ is employed with code block of $1152$ information bits. The log-MAP BCJR algorithm has been employed for SISO channel decoding.

In Fig. 1 the BER performance of the full proposed scheme (including both scenario-adaptive detection and decoding) is depicted for several TER values ($10^{-4}$, $10^{-3}$, $10^{-2}$) compared to the typical. As targeted, significant performance degradation is observed only for those signal-to-noise (SNR) values which provide BER performance better than the TER. In addition, small, unwanted performance degradation can be observed before reaching the TER due to the approximate nature of (9).

In Fig. 2 the performance of the proposed scheme with only scenario-adaptive detection and full channel decoding is depicted. It is shown that the proposed LLR clipping can preserve the optimal performance for SNR ranges up to the one providing the TER one.

In Fig. 3 the complexity of the SD is depicted in terms of average visited nodes for the several TER values. It is shown that significant complexity gains can be achieved over the whole SNR range (even when TER has not still be reached) which shows the high efficiency of the LLR clipping. For SNR=14 dB and TER=$10^{-4}$ the gain in comparison with the typical solution (i.e., without clipping) reaches 92%. Finally, an additional gain from 28-36% can be observed for any additional TER increase of one order of magnitude. In Fig. 4 the average required $\beta$ stores are depicted. A reduction of 37% is observed for SNR=14 dB and TER=$10^{-4}$, while a gain up to 35% (for SNR=14 dB) can be observed for any TER increase of one order of magnitude.

As already discussed in Section I, the proposed method does not target the minimum SD complexity since such an approach would include additional burden which would decrease the practicality and the applicability of the scheme (see Section III, step 2). However, it exploits the additional (from the minimum) processing overhead to decrease the power consumption of the channel decoder. In this context, it would be of interest to give a hint on the complexity gains originating from those two approaches, even after ignoring the burden and the additional processing cost of the required performance prediction. From Fig. 2 it can be seen that if prediction methods were available they would select an LLR clipping value of $\ln\left(1/10^{-2}-1\right)$ for SNR=12 dB and a TER of $10^{-3}$, while the proposed approach employs an LLR clipping value of $\ln\left(1/10^{-3}-1\right)$. This results in a complexity overhead of about 23% (see Fig. 3) at the SD side. However, in contrast to the minimum SD complexity approach, the proposed one allows a reduction of 25% on the number of required $\beta$ stores. In the same way, at 13.7 dB and for TER=$10^{-4}$ the optimal LLR clipping value would be again $\ln\left(1/10^{-2}-1\right)$. Then, the proposed approach would result in an overhead of 48%, but it would allow a reduction of 51% on the number of the required $\beta$ stores.

## V. CONCLUSION

A practical and broadly applicable MIMO-APP receiver processing framework has been proposed which allows the adjustment of the receiver processing requirements (i.e., of the soft-output detector and the SISO channel decoder) to the transmission conditions and the required BER. In contrast to receivers supporting multiple detection schemes, the proposed approach does not require error prediction and it employs one single soft-output depth-first sphere decoder (SD) able to adjust its complexity by means of (BER) *performance-driven* LLR clipping. Since the proposed approach *does not* target the minimum detector's complexity the corresponding processing overhead is exploited to reduce the energy consumption of the SISO channel decoder. Despite the simplicity and the easy applicability of the approach, significant complexity savings can be observed both at the soft-output detector and the channel decoder.

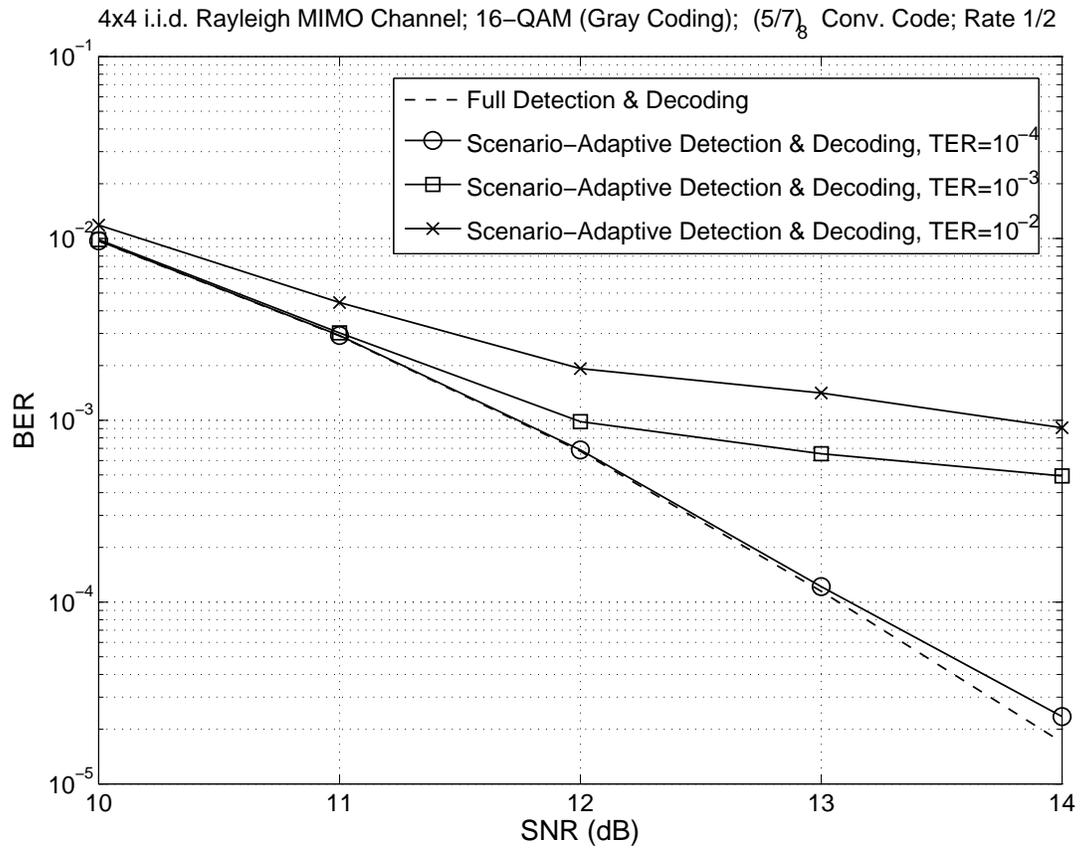

Fig. 1.   BER performance with scenario-adaptive detection and decoding for several TER values.

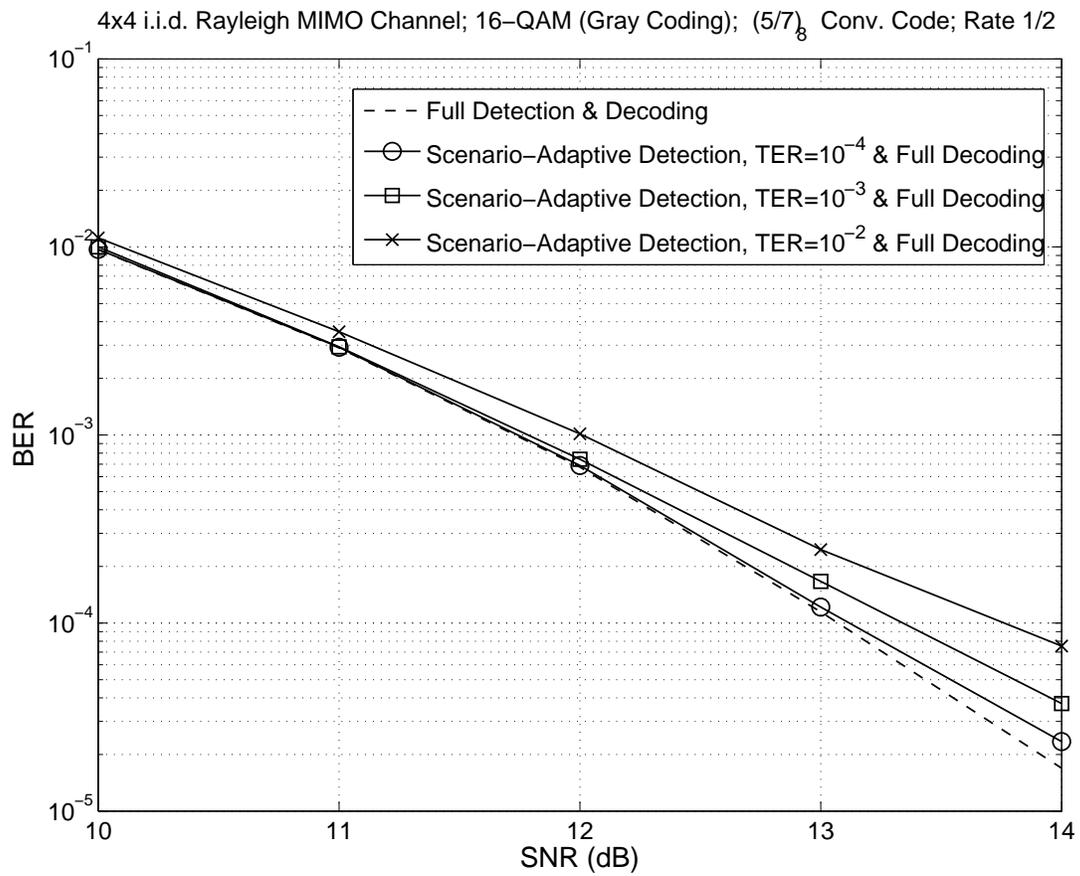

Fig. 2.   BER performance with scenario-adaptive detection and full decoding for several TER values.

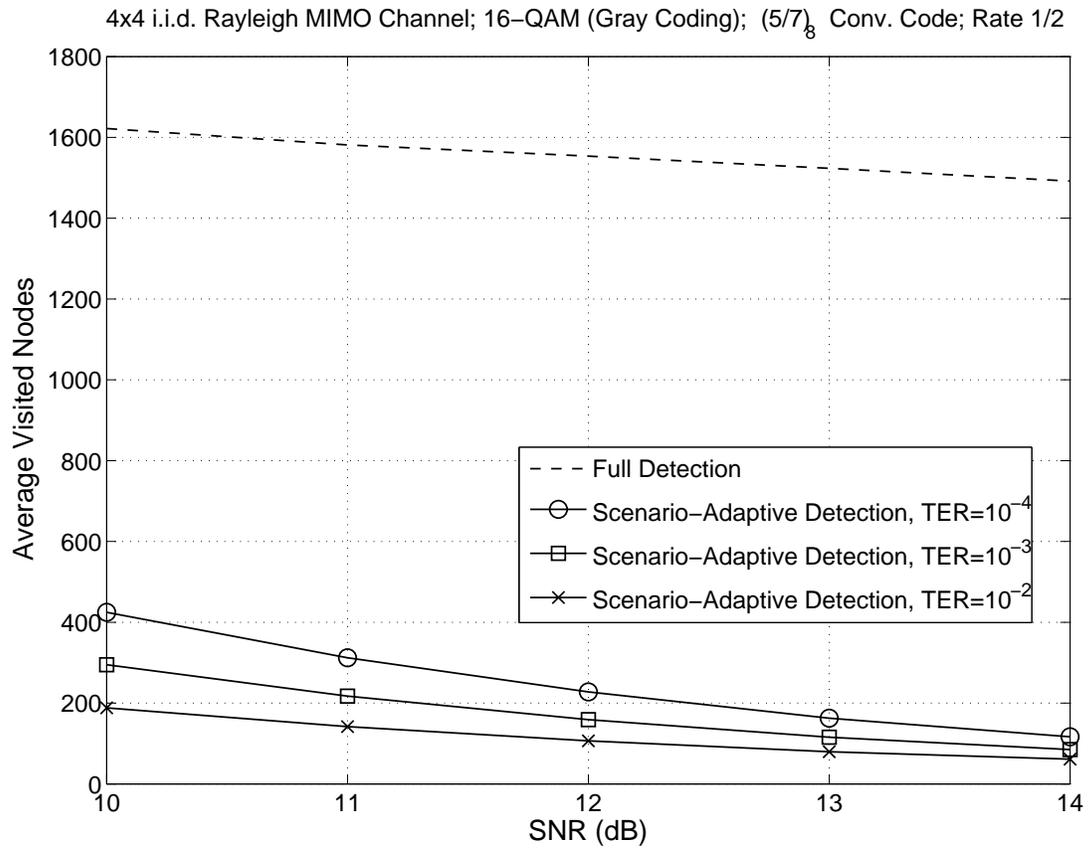

Fig. 3. Complexity of scenario-adaptive soft-detection in terms of average visited nodes for several TER values.

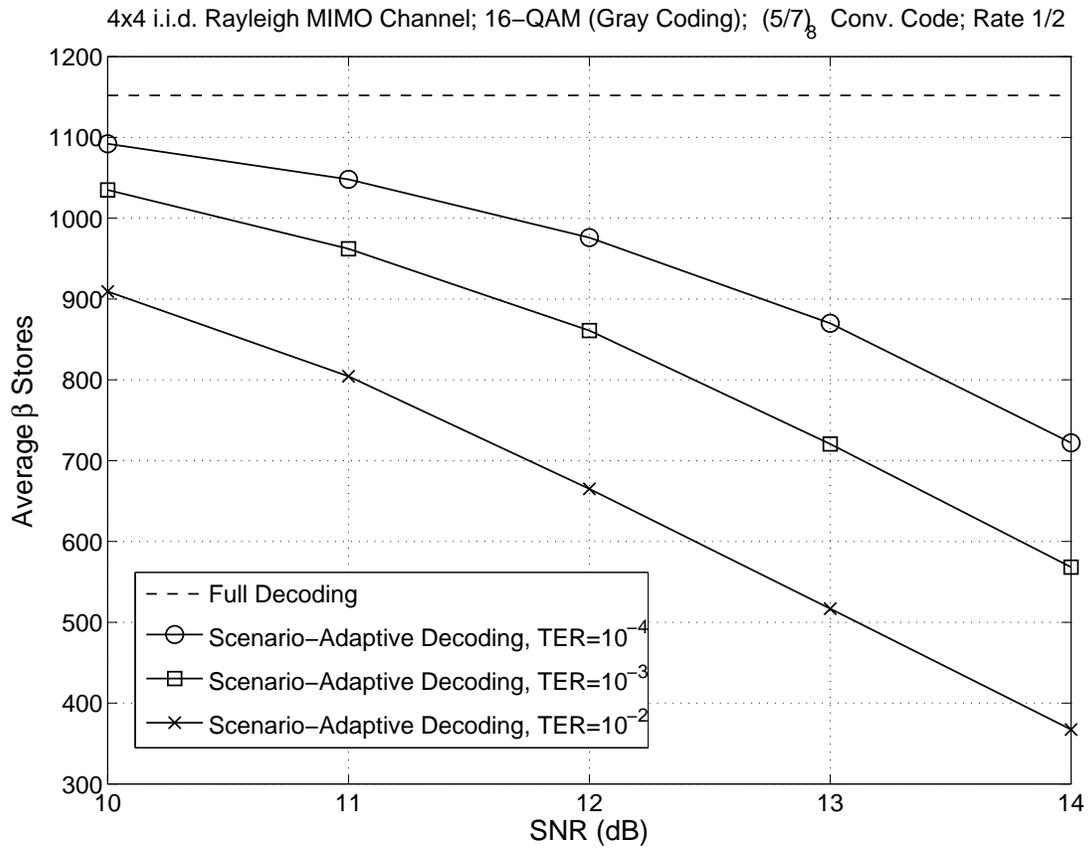

Fig. 4. Average required $\beta$ stores for scenario-adaptive SISO channel decoding and several TER values.